# Current progress in corrosion of multi principal element alloys

M. Ghorbani[1], Z. Li[2], Y. Qiu[1,3], P. Marcus[4], J.R. Scully[5], O. Gharbi[6], H. Luo[7], R.K. Gupta[8], Z.R. Zeng[9], H.L. Fraser[10], M.L. Taheri[11] and N. Birbilis[1,*]

[1]Faculty of Science, Engineering and the Built Environment, Deakin University, VIC., Australia.
[2]College of Engineering, Computing and Cybernetics, Australian National University, A.C.T., Australia.
[3]Faculty of Materials, Wuhan University of Science and Technology, Wuhan, PR China
[4]PSL University, CNRS, Chimie ParisTech, Institut de Recherche de Chimie Paris, Physical Chemistry of Surfaces Group, Paris, France
[5]Department of Materials Science and Engineering, University of Virginia, VA, USA
[6]CNRS, Sorbonne Université, Laboratoire de Réactivité de Surface, LRS, Paris, France
[7]National Materials Corrosion and Protection Data Center, Institute for Advanced Materials and Technology, University of Science and Technology Beijing, Beijing, PR China
[8]Department of Materials Science and Engineering, North Carolina State University, Raleigh, NC, USA.
[9]State Key Laboratory of Advanced Design and Manufacturing Technology for Vehicle, Hunan University, Changsha, China.
[10]Department of Materials Science and Engineering, The Ohio State University, Columbus, OH, USA.
[11]Department of Materials Science and Engineering, Johns Hopkins University, Baltimore, MD, USA.

*nick.birbilis@deakin.edu.au

**Abstract**

Whilst multi-principal element alloys (MPEAs) remain a promising class of materials owing to several attractive mechanical properties, their corrosion performance is also unique. In this concise review, we present an emerging overview of some of the general features related to MPEA corrosion, following a decade of work in the field. This includes highlighting some of the key aspects related to the electrochemical phenomena in MPEA corrosion, and the relevant future works required for a holistic mechanistic understanding. In addition, a comprehensive database of the reported corrosion performance of MPEAs is presented – based on works reported to date. The database is assembled to also allow users to undertake machine learning or their own data analysis, with a parsed representation of alloy composition, test electrolyte, and corrosion related parameters.

# 1. Introduction

The corrosion of multi-principal element alloys (MPEAs) remains an active area of research, increasing in intensity over the past five years – largely owing to the interest afforded to the so-called high entropy alloys (HEAs) [1]. Herein, MPEAs is a broad terminology used to describe alloys containing at least two principal alloying elements, inclusive of HEAs, compositionally complex alloys and complex concentrated alloys [2-4].

Of the growing reports regarding MPEAs to date, it is evident that unique microstructures and elemental combinations possible for MPEAs, also contribute to unique mechanical properties. Such properties may be favourable, and can include high hardness and strength, high thermal stability, along with reports of excellent corrosion resistance [5-7] The potential (industrial) use of emerging MPEAs would appear imminent for niche applications [8] and correspondingly, a detailed understanding of the key mechanistic facets of MPEAs corrosion is critical.

From a corrosion perspective, there are comparatively fewer reports regarding MPEAs with respect to reports of microstructures and mechanical properties. An early comprehensive-type review exploring the data available to date was reported by Tang and co-workers in 2014 [9]. Another early synopsis was provided by Qiu and co-workers in 2017 [6], who summarised that generally speaking, a variety of HEAs performed similarly to so-called CRAs (corrosion resistant alloys) of the stainless-steel families, whilst also identifying that test electrolyte plays a key role in the assessment of corrosion performance. A later attempt at a mechanistic interpretation of corrosion performance of MPEAs was proposed by a selection of the present authors [1], where a review of the literature to date identified unique characteristics of the passive films formed upon MPEAs – a theme that is reinforced further herein from more recent (and diverse) independent studies that have employed detailed surface characterisation including x-ray photoelectron spectroscopy.

Owing to the volume of emerging studies regarding the corrosion of MPEAs, there is already an emerging nuance in corrosion mechanisms, which is to be expected. This is because MPEAs cover such a diverse range of alloy compositions (often completely diverse such that there is little to no elemental overlap between different alloys classed as MPEAs) that generalisation of corrosion performance is not possible. As a consequence, rather than provide an incremental review of the field herein, an alternate approach is taken whereby a focus on 'data' presentation is the primary aim. Specifically, the curation of the most comprehensive database of corrosion properties reported to date for MPEAs is compiled, and presented. A database of corrosion properties for MPEAs may serve as a valuable resource – from enabling alloy screening, and the optimisation (of sub-compositions) through data science approaches. This may also help reduce the trial-and-error approach in alloy development – as noted in emerging databases for mechanical properties of MPEAs [10, 11]. There is also little doubt that as the volume of data regarding MPEA properties continues to increase, that the utilisation of machine learning (ML) will be indispensable in effectively handling and analysing complex datasets of high dimensionality [12]. A recent example of the use of machine learning in the selection and prediction of corrosion in multi-principal alloys was presented by Roy and co-workers [13] - who summarised that while it is possible to down select corrosion-resistant MPEAs by using ML from a large search space - a larger dataset (inclusive of higher quality data) is needed to accurately predict the corrosion rate of MPEAs. To this end, the presentation of a comprehensive database of corrosion properties is warranted. In addition, the careful (human expert) assessment of data quality in the reported literature is also essential in order to avoid spurious data. The remainder of this concise review will identify some key mechanistic insights, present a unique database, and also highlight prospects for future work.

## 2. A selection of mechanistic facets related to MPEA corrosion

### 2.1 Incongruent dissolution

Many corrosion studies of MPEAs (inclusive of HEAs) report relatively low rates of corrosion. Therefore, in such cases, it is nowadays customary to accompany the corrosion testing with some surface analysis or microscopy - to 'see' and visualise the nature of the corrosion upon the MPEA surface. A typical characteristic of MPEA corrosion is what is defined as 'incongruent' dissolution – which implies a non-uniform dissolution of the alloy during corrosion. In fact, arguably since the widespread uptake of field emission gun scanning electron microscopy in corrosion research – the corrosion (specifically, dissolution) of most engineering alloys is accompanied by incongruent dissolution [14]. This implies that one or more elements within the alloy under investigation will preferentially dissolve. Such incongruent dissolution is identified by either a selective leaching from, or accumulation of elements upon, the alloy surface [2]. Owing to the 'cocktail' of elements typical in MPEAs (and HEAs), incongruent dissolution is a characteristic of essentially all alloys investigated to date, evidenced in cases where the assessment of corrosion has been sensitive to assessment of this phenomena. The extent of incongruent dissolution has been thoroughly and quantitatively characterised in cases where the ASEC (atomic spectro-electrochemistry) has been utilised. The ASEC method allows assessment of dissolution in an element-by-element manner, during polarisation or open circuit exposure. Examples include the following studies:

**$Al_{0.3}Cr_{0.5}Fe_2Mo_xNi_{1.5}Ti_{0.3}$.** In a recent study by Inman and co-workers [15], the AlCrFeMoNiTi system was explored with various Mo concentrations using ASEC. The fidelity (and resolution) of the approach is depicted in **Figure 1**, where it is revealed that over a range of potentials, the rate of elemental dissolution varies significantly on an element-by-element basis. The rate of dissolution of Fe and Ni is markedly greater than the rate of dissolution of Cr and Al, whilst minimal dissolution of Ti and Mo was observed. Similar results were also reported upon the $Al_{0.3}Cr_{0.5}Fe_2Mn_{0.25}Mo_{0.15}Ni_{1.5}Ti_{0.3}$ alloy [16].

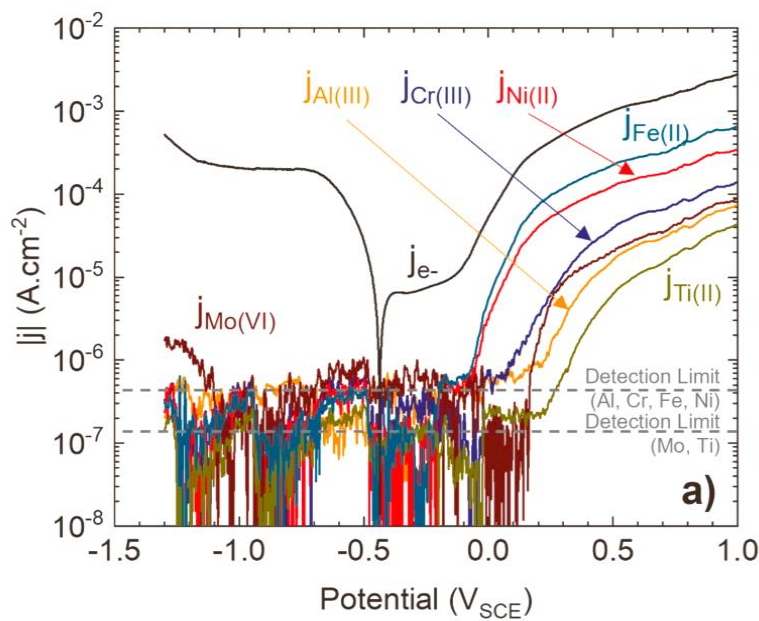

*Figure 1. Equivalent current densities of in-situ elemental dissolution rates monitored via ASEC during polarization of the base alloy of $Al_{0.3}Cr_{0.5}Fe_2Mo_xNi_{1.5}Ti_{0.3}$ with 3.2 at. % Mo in nitrogen sparged 0.1 M NaCl (pH 4). Dashed lines indicate element detection limits.*

**$Ni_{38}Fe_{20}Cr_xMn_{21-0.5x}Co_{21-0.5x}$.** In a comprehensive study by Gerard and co-workers (which also involved detailed morphological characterisation), it was noted for the NiFeCrMnCo system that all of Fe, Co, and Mn selectively depleted in the tests conditions explored [17]. Nickel was found to enrich at the film metal interface forming an altered zone. The study was carried out in dilute NaCl electrolyte, with pH adjusted to 4 using 0.1 M HCl. The result of incongruent dissolution was that passive current density was correlated with the Cr cation fraction in the passive film – with Cr enrichment in the passive film dependent on alloy composition.

**AlTiCrV**. The equiatomic four-component MPEA has been widely study by Qiu and Choudhary [18-20] owing to a rather exceptional corrosion resistance in this alloy (and the related family of alloys [21]). It was revealed using ASEC that the alloy dissolution was incongruent, occurring in a dissolution profile in the order of Cr > Al >> V >> Ti. The extent of incongruent dissolution is very significant in this alloy, akin to the well documented de-alloying of several single-principal element alloys (such as brasses) [22], albeit that the alloy is an essentially 'perfectly uniform' single phase alloy in regards to elemental distribution – as confirmed by atom probe tomography [23]. Whilst the extent of incongruent dissolution is significant (i.e. comparatively speaking, one element dissolves at greater than an order of magnitude in relation to another element), it is noted that the total extent of dissolution for AlTiCrV remains 'very low', in that the dissolution (and corrosion) rate of the alloy are lower than those of a wide variety of stainless steels and MPEAs in general.

**AlFeMnSi**. This equiatomic MPEA developed by O'Brien and Gupta has an absence of what may be deemed conventionally 'corrosion resistant elements' such as Cr, V, Mo, etc. [24, 25]. However, the alloy presents excellent corrosion resistance – including low rates of dissolution across a range of NaCl containing electrolytes and potentials. The alloy dissolution was studied by ASEC, and it was also revealed that albeit dissolution rates are overall low, the dissolution proportions occur in the following order Si > Al» Fe > Mn.

**CoCrFeMnNi.** This alloy, (known as the Cantor alloy) was studied by Li and co-workers using ASEC [26], however the alloy explored was a variant of the more common HEA and included 0.52 at. % nitrogen (N). Whilst the study covered a range of facets, including exploration of N on passivation kinetics, it was noted that incongruent dissolution was dominant. In the case of the CoCrFeMnNi alloy, either with or without N, incongruent dissolution was clearly observed for Cr, and to a lesser extent, for Mn. In another study on another equiatomic MPEA alloy that also contains Cr, Fe, and Mn, the incongruent dissolution of **CrFeMnV** was also noted by O'Brien and co-workers – whereby Cr oxidation was dominant [27].

The above emphasis on incongruent dissolution, albeit concise, is important on the basis that the corrosion properties of MPEAs have been reported to alter (with some significance) as a result of variations in proportion to the ratio of alloying additions. In such cases, the relative proportions of incongruent dissolution also vary (and, in turn, dominate corrosion kinetics). An example (amongst many) is in the work exploring $Al_x(CoCrFeNi)_{100-x}$ (where x was varied between 0 and 20). A critical aspect of MPEAs compositional changes is also the alteration of alloy microstructure. For instance, nanostructural ordering in a single-phase alloy may effect corrosion [28]; whilst alloys may also transition from single phase (i.e., FCC, or BCC, or HCP), to dual phase (e.g. FCC+BCC), to multi-phase (inclusive of phases that could include B2, Laves, sigma, or diverse intermetallics). Such microstructural evolution has been documented in some discrete corrosion studies, such as the work of Muangton and co-workers exploring the CoCrFeNi-x (x = Cu, Al, Sn) system [29], and the work of Shi et. al. [30] who explored the variation in corrosion with Al content for Al$_x$CoCrFeNi HEAs. It has also been established that Mn is disruptive to passivation, but a good stabilizer of FCC nanostructures in NiCrFeCoNiMn [31]. It was recently reported that Mn may be optimised at around ~5% to promote the FCC phase, whilst less than ~15% is essential to maintain corrosion resistance in AlCrTiFeNiMnMo MPEAs [32].

The review herein – as a general review - will not focus on the specific facets of microstructural effects on corrosion of MPEAs for three reasons: (i) the prevalence of incongruent dissolution (and its subsequent effect on passivity as outlined below) would appear, at present, to be the dominant mechanism controlling corrosion of MPEAs; (ii) the alloy specific details regarding microstructural effects are best interpreted in depth from each associated study (which may be accessed from the accompanying reference list within the MPEA corrosion database); and, (iii) detailed microstructural studies are typically covered in the domain works associated with the mechanical properties of MPEAs.

### 2.2 'Passive' films upon MPEAs

Since ~2020, the number of works that have studied passive films upon MPEAs has increased significantly, relative to all works prior to 2020. To date, the principal tool for studying passivity of MPEAs in a physical sense, is x-ray photoelectrochemical spectroscopy (XPS). When XPS is conducted in concert with attendant/associated electrochemical testing (be it potentiodynamic polarization or electrochemical impedance spectroscopy) XPS can provide significant insights regarding the structure and development of passive films upon MPEAs. This has been observed for studies exploring AlCoCrFeNi [33, 34],CoCrFeMnNi [35-37], FeCrNiCoMo [38], CrMnNiFe [39], TaNbHfZrTi [40], AlCrFeMoV [41], AlCoCrFeNiTi [42], TiTaHf(Nb, Zr) [43], AlNiCoFeCu(Cr) [44], CoCrFeNiZr [45], AlCoFeNiTiZr [46], NiCrFeRuMoW [47], TiZrHfNbTa [48], AlNbTiZrSi [49], FeMnCoCrSiAl [50], AlCoCrCuFeNiB [51], VAlTiCrSi [52], along with XPS having also been carried out on the above aforementioned AlCrFeMoNiTi, Al$_x$TiCrV, AlFeMnSi and CrFeMnV alloys. The present authors note that there is considerable insight and meticulous work in the studies cited as having conducted XPS upon MPEAs. Each provides a unique insight, and includes several aspects of what are deterministic interpretations of MPEA passivity. The first is that oxides may contain cations of many, but not all of, the alloying elements found in the substrate - in aerated aqueous environments. Another of the findings evident of the XPS studies, is that in many cases, there was strong evidence of unoxidized metal detected in the surface films. In these cases the unoxidized metal in the zero (0) oxidation state, M$^0$, was interpretated to not arise as a result of photoelectrons from the underlying metallic substrate (using grazing incidence and low energy XPS). The presence of unoxidized metal in the surface (passive) films upon MPEAs may be a unique feature, compared with most traditional single principal element alloys. Unoxidized metal was also observed for alloys of vastly differing composition [53]. The presence of such unoxidized metal was posited, coincident with the observation of its prevalence, to likely play a key role in the corrosion resistance of MPEAs – owing to the notion that unoxidized metal may help accommodate strain within surface films [18]. An example of XPS spectra collected upon an FeCoCrNiAl alloy (following exposure to 0.5 M H$_2$SO$_4$) are presented in **Figure 2**, whereby the peaks associated with M$^0$ peaks have been labelled/annotated.

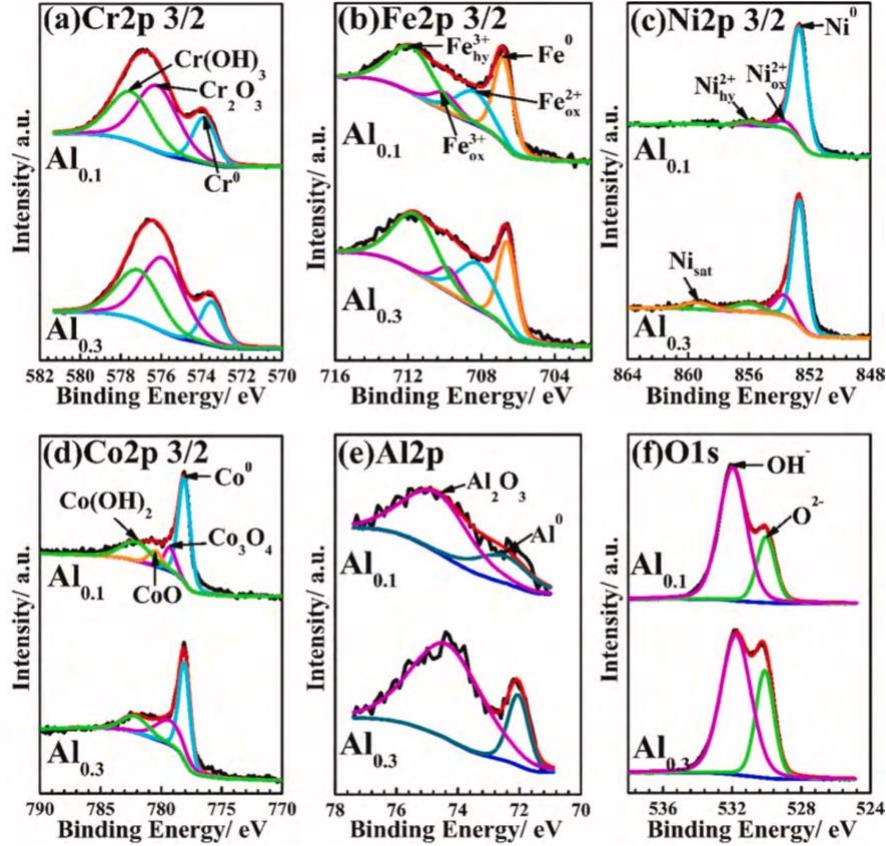

*Figure 2. XPS results of O 1s, Cr 2p3/2, Fe 2p3/2, Co 2p3/2, Ni 2p3/2 and Al 2p for the surface film formed upon following exposure to 0.5 M $H_2SO_4$ – for $FeCoCrNiAl_x$ (where x = 0.1 or 0.3, as depicted in the figure) [34]*

The relative abundance of unoxidized metal present in surface/passive films upon MPEAs varies across the studies cited above, however is considered to be appreciable. An example from the equi-atomic MPEA, CrFeMnV, is presented in **Figure 3**.

From Figure 3, it is evident that for the surface film formed upon CrFeMnV following 1hr immersion in 0.6M NaCl, the ratio of unoxidized to oxidized species is ~14% for Cr, ~69% for Fe, ~41% for Mn, and ~20% for V. These are appreciable proportions, and also highlight another critical factor related to passive films upon MPEAs. Namely, not only is the proportion of unoxidized to oxidized species for individual elements different (depending on the element type, and the alloy they occupy), but the composition of the surface film also highly deviates from the stoichiometric composition of the base MPEA. In the case of the equiatomic base CrFeMnV, the surface film composition is indicative of rather highly incongruent dissolution, which leads to a very unique surface film composition (and proportion of elements, oxides, hydroxides, that occupy the surface film). It is notable that Cr(III) cation fraction is much greater than its alloy concentration in the equi-atomic alloy, while Mn and Fe are present in small concentrations [47].

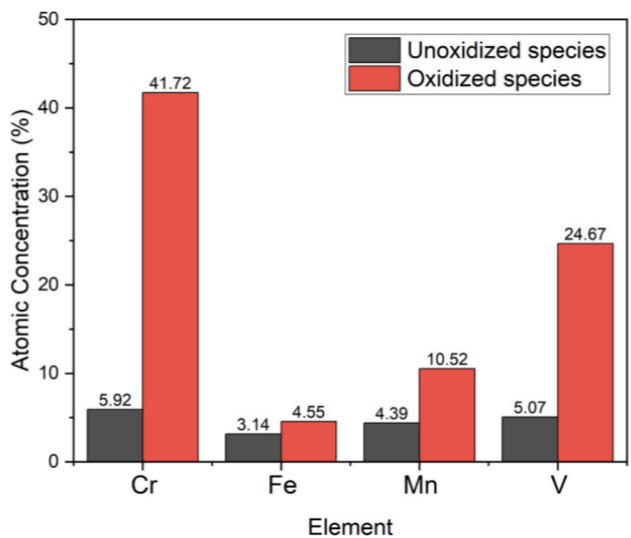

*Figure 3. Atomic concentration of unoxidized and oxidized species on the surface of CrFeMnV as determined using XPS, after 1 h immersion in 0.6 M NaCl. [27]*

The complexity of surface/passive films for MPEAs that is embodied in the results of **Figure 3** (and that arise from incongruent dissolution) is also far more intricate than may be first considered. This is because minor changes in MPEA composition will impact the extent of incongruent dissolution, which will have a direct impact on the proportion of elements (and their compounds) found upon the surface of an MPEA. For example, in some MPEAs, the incongruent dissolution of Cr may dominate the dissolution profile, whilst in other Cr-containing MPEAs, it may not necessarily be the principal element dissolving. Another important nuance is that MPEAs also have rather complex surface film compositions (by compound type) *and* morphologies. Using another equiatomic alloy as an example (equiatomic alloys are useful as examples on the basis that a deviation from stoichiometric dissolution or surface films highlights complexity) – is the CoCrFeMnNi alloy studied by Wang and co-workers [35]. The study by Wang is interesting for two reasons. The first is that it is one of only very few studies to have complemented XPS with time-of-flight secondary ion mass spectrometry (ToF-SIMS), and the second is that it has highlighted aspects of surface film morphology. There was strong correlation between XPS and ToF-SIMS, whilst the latter method was also indispensable in illuminating depth specific facets. For CoCrFeMnNi, there was a so-called duplex film, comprising a principally Cr/Mn inner oxide layer, and an Cr/Fe/Co outer oxide/hydroxide layer. It was also noted that the passive film was remarkably thin (consistent with other studies of MPEAs) at ~1.6 nm, and the native oxide (from air exposure) was also thin (~1.4 nm) but was also a duplex film. This latter point is an exemplar that typifies what is generally observed from MPEAs, in that even in air, rapid oxidation occurs, which is accompanied by subsequent passivity (and high corrosion resistance). There is an irony at the superficial level, whereby the high corrosion resistance of MPEAs is causally associated with very high rates of 'reactivity' that leads to dynamically formed (and highly stable) surface films.

Following the above context, there is therefore a variety of mechanistic reasons reported in the literature (which are MPEA dependent) for the high levels of corrosion resistance and passivity. The reasons for high corrosion resistance of MPEAs are diverse, and each is justified according to the unique circumstances by which each MPEA incongruently dissolves. Even in an extensive review, it is not possible to cover each mechanism, however an attempt is made to highlight a few diverse reasons for high corrosion resistance in some MPEAs, with an emphasis on equiatomic alloys as examples.

**CoCrFeMnNi**. The transformation from Cr hydroxide to Cr oxide (by dehydroxylation) was observed for increasing passivation time. This is accompanied by a significant enrichment in Ni directly beneath the surface film [35].

**AlFeMnSi**. A significant extent of incongruent dissolution was observed, that corresponded to ~70% of dissolution current being Si dissolution. The dissolved Si was redeposited at the alloy surface, and was rapidly 'dynamically' precipitated on the outer surface of the alloy as a Si-hydroxide surface film. This surface film suppressed further dissolution and resulted in excellent passivity of the alloy [25].

**CoCrFeNi**. Passive film thickness increases with increasing pH and exposure duration [54]. XPS revealed that the fraction of Co(II) and Ni(II) oxides in the passive film were important contributors to the passivity of the alloy (as assessed from variations in electrolyte pH). However, the passive film was a multi-oxide passive film, that was highly enriched with stable $Cr_2O_3$.

**AlTiCrV**. Whilst this alloy was previously discussed, the example is elaborated here, because it is one of the studies in which ASEC has been utilised (along with XPS), and where repassivation has also been studied – all in NaCl environments. The passivation (and exceptional re-passivation) mechanism for this alloy is embodied in **Figure 4**.

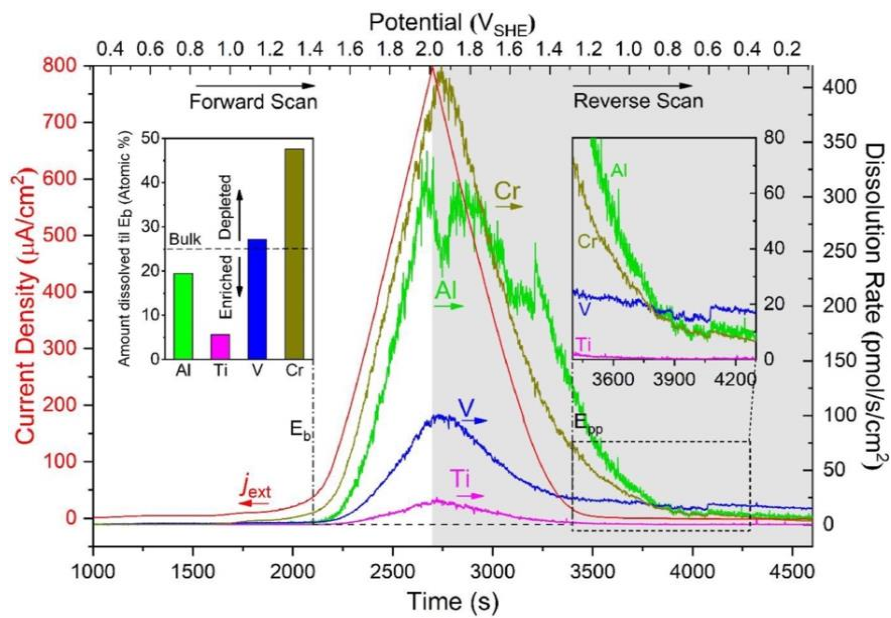

*Figure 4. ASEC cyclic polarisation curve of AlTiVCr showing incongruent dissolution in dilute NaCl, during a forward and reverse potential scan. Significant enrichment of Ti on the alloy surface occurs, resulting in the rapid repassivation during the reverse potential scan. $E_b$ and $E_{pp}$ represent final breakdown potential and protection potential, respectively [19].*

From **Figure 4** it is evident that incongruent dissolution during anodic polarisation is associated with the dissolution current being principally associated with Cr and Al dissolution. Concomitantly, there is significant enrichment of Ti during passivation (principally as the highly stable $TiO_2$). In turn, when the alloy is reverse-polarised (following transpassive polarisation), the alloy surface is different to that prior to anodic polarisation. The alloy surface is highly enriched in stable $TiO_2$, providing a very rapid 're-passivation' and enduring passivity. Again, this highlights the unique nature of passivity for each MPEA explored in detail through XPS and ancillary analytical methods.

In non equiatomic MPEAs, there are also diverse mechanisms for passivity reported (fittingly for the alloys under investigation) such as the "unexpected" high Cr enrichment within the passive film during the aqueous passivation of multi-principal element alloy $Ni_{38}Fe_{20}Cr_{22}Mn_{10}Co_{10}$ [2]. In a follow-on study, the same authors were able to really highlight the complexity of MPEA surface films including the possibility of the presence of Cr(III) oxides, hydroxides, and complex oxides such as possible spinels ($FeCr_2O_4$ and $NiCr_2O_4$), which could not be ruled out [17].

Additional works that have sought to provide mechanistic insights include the work of Bi and co-workers, who very recently studied $AlNbTiZrSi_x$ alloys [49], and the work of Fu et. al. who studied Al-containing MPEAs [34]. The latter study explored $FeCoCrNiAl_x$, with varying Al concentrations revealing an increase in Al content was associated with enhancement of the oxide/hydroxide ratios for Cr, Fe and Al, and improved passivation (from electrochemical assessment)

It is certainly conceded that whilst the above (concise) treatise may attempt to cover key facets of MPEA corrosion and passivity to date, there remains significant knowledge gaps. In this review, facets of repassivation of MPEAs have not been covered in detail (as that would merit a separate electrochemical treatise), and similarly aspects of 'passivation models' that have been under evolution in the field for now over a century, also merit a separate treatise. However, it is noted that a very recent study by Wang, Marcus and co-workers [55] has provided the prospect to enhance the collective understanding of passivation in complex MPEAs. In that work, ToF-SIMS was combined with XPS and deuterium labelling ($D_2O$), in order to investigate the interfacial transport mechanisms of hydroxyls between the aqueous electrolyte and the surface film during passivation of the $Cr_{15}Fe_{10}Co_5Ni_{60}Mo_{10}$ MPEA. This study was able to provide an enhanced insight into the mechanisms of passive film growth and during anodic polarisation (i.e. during passivation) and may hold an important place in benchmarking emerging models of 'passivity' that can harness the complexity of MPEAs. In another recent work from the same group [56] the mechanisms of the enhancement of passivity of MPEAs containing chromium and molybdenum, whereby Mo(IV) oxide species are concentrated close to the CrIII-rich inner barrier layer, were elucidated.

## 3. A survey of MPEA corrosion

Whilst the complexity of corrosion (and passivation) of MPEAs is undoubtedly a mechanistically complex endeavour, the review herein seeks to provide a holistic treatise of the field by providing a unique set of data that is useful to the field as a compendium, and with a data science lens. To this end, a synopsis of the corrosion associated data for MPEAs is provided.

### 3.1 Database of corrosion properties

The development of a database of corrosion properties for MPEAs was carried out by eight individuals over a period of several months. This included manual trawling of journal articles and archival literature. The requirement for manual trawling was that much of the data collected was extracted from plots or non-text representations, to allow an expert level assessment of whether data was valid (or spurious, and therefore to be omitted, for which there are many examples [57, 58]). The database of MPEA corrosion properties assembled for this review is open-access, can be retrieved at https://data.mendeley.com/datasets/nskb9khgsm/1 [59].

Characteristics of the MPEA corrosion database are presented in **Table 1**, indicating the types of alloying elements present in the alloys within the database, their corresponding range, the mean value in the alloys they populate, and the number of times each element is present in the database (with a total of 619 unique alloy entries).

*Table 1. Characteristics of data in MPEA corrosion database, analysed with respect to individual alloying elements present in MPEAs*

| Element | Range as fraction of alloy (atomic fraction) | Mean | Count |
|---|---|---|---|
| Fe | 0.0 - 0.53 | 0.23 | 510 |
| Ni | 0.0 - 0.75 | 0.22 | 485 |
| Cr | 0.0 - 0.333 | 0.19 | 481 |
| Co | 0.0 - 0.5 | 0.2 | 395 |
| Al | 0.0 - 0.69 | 0.15 | 353 |
| Cu | 0.0 - 0.25 | 0.15 | 229 |
| Ti | 0.0 - 0.8 | 0.2 | 199 |
| Nb | 0.0 - 0.333 | 0.16 | 86 |
| Mn | 0.0 - 0.3 | 0.2 | 84 |
| Mo | 0.0 - 0.615 | 0.13 | 84 |
| Zr | 0.0 - 0.37 | 0.18 | 77 |
| V | 0.0 - 0.364 | 0.2 | 65 |
| Si | 0.0 - 0.2 | 0.1 | 21 |
| W | 0.0 - 0.2 | 0.16 | 21 |
| Ta | 0.0 - 0.25 | 0.18 | 18 |
| Hf | 0.0 - 0.278 | 0.2 | 17 |
| Sn | 0.0 - 0.2 | 0.03 | 14 |
| B | 0.0 - 0.154 | 0.05 | 12 |
| Y | 0.0 - 0.2 | 0.2 | 6 |
| Mg | 0.0 - 0.25 | 0.21 | 3 |
| Zn | 0.0 - 0.3 | 0.3 | 3 |
| La | 0.0 - 0.024 | 0.02 | 1 |
| Ga | 0.0 - 0.25 | 0.25 | 1 |
| Be | 0.0 - 0.25 | 0.25 | 1 |

The MPEA corrosion database seeks to provide a holistic treatise of reported corrosion data for MPEAs. As such, the database includes entries for alloys that have been tested in a variety of electrolytes, of differing electrolyte concentration. The associated data characteristics related to test electrolytes are provided in **Table 2**., and the data characteristics related to the reported corrosion related parameters are provided in **Table 3**.

*Table 2. Characteristics of data in MPEA corrosion database, presented with respect to test electrolytes reported.*

| Electrolyte type | Range (Concentration in M) | Count |
| --- | --- | --- |
| NaCl | 0.1 - 5 | 372 |
| $H_2SO_4$ | 0.05 - 1.088 | 105 |
| Sea water | 0.055 - 5.5 | 38 |
| $HNO_3$ | 0.5 - 2.38 | 36 |
| NaOH | 1 - 5 | 26 |
| PBS | n/a | 13 |
| HCl | 0.5 - 6 | 8 |
| Hanks | n/a | 5 |
| KOH | 1 - 1 | 4 |

*Table 3. Characteristics of data in MPEA corrosion database, presented with respect to the range of values for corrosion related parameters*

| Parameter | Range | Count |
| --- | --- | --- |
| Corrosion potential (mV vs. SCE) | -1460 - +290 | 604 |
| $i_{corr}$ ($\mu$A/cm$^2$) | 0.0003 - 5080 | 580 |
| Pitting potential (mV vs. SCE) | -949 - +2311 | 329 |
| Calculated passive window (mV) | -249 - +2141 | 343 |

The notation in Table 3, and in the MPEA corrosion database, utilizes the descriptor of 'corrosion potential'. It is noted that in some studies, open circuit potential (from the open circuit measurement of electrode potential) may have been reported, however in the vast majority of cases it was not reported – rather the $E_{corr}$ value was reported. The $E_{corr}$ value is determined during polarisation testing, and may slightly differ from the corrosion potential (although this variation is typically within a small range (of mV), and in cases where testing has been well executed, there is alignment between corrosion potential and $E_{corr}$. In the present study, it was decided to use the singular notation of 'corrosion potential', to embody the presentation of the parameter that is either the open circuit potential or the $E_{corr}$ value from each study evaluated.

The distribution of corrosion related parameters of alloys in the MPEA corrosion database is presented in **Figure 5**. The distributions indicate rather vividly that the distribution mode (where the mode is defined as the most frequently occurring score, corresponding to the highest point on the distribution plot) reveals: corrosion potential values of ~ -280 mV$_{SCE}$, corrosion current density values $<< 0.5 \mu$A/cm$^2$, and pitting potential values of ~ +180 mV$_{SCE}$.

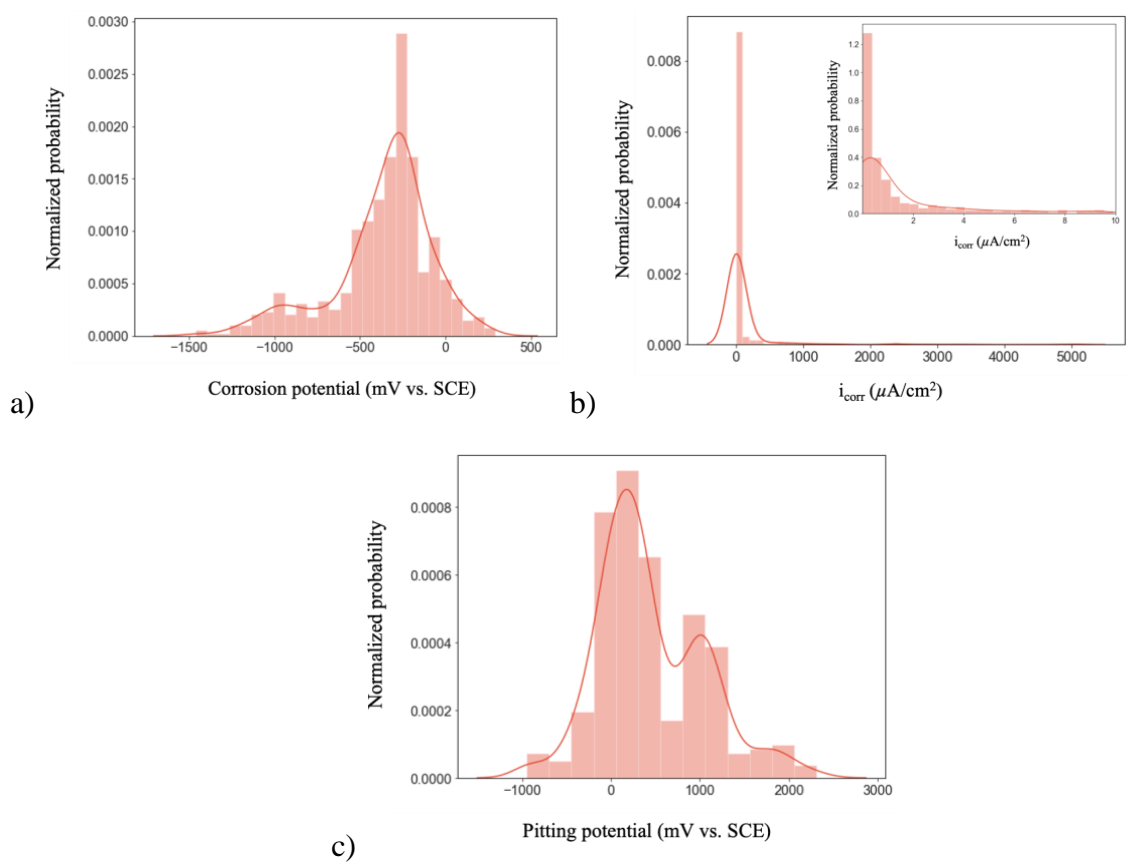

*Figure 5.* Data distributions for corrosion related parameters of alloys in the MPEA corrosion database, showing a) corrosion potential, b) corrosion current density, and c) pitting potential.

### 3.2 Characteristics of MPEA corrosion data

*3.2.1. Overall data trends*

To better visualise the data embodied in the MPEA corrosion database, a number of characteristic plots have been generated, to show multiple parameters at once, with data disambiguated by test electrolyte. The corrosion potential data versus the corresponding $i_{corr}$ is presented in **Figure 6a**, disambiguated by test electrolyte. Similarly, pitting potential and calculated passive window are also shown in **Figure 6b** and **6c**, respectively. The calculated passive window is equal to: (pitting potential) – (corrosion potential). The calculated passive window could only be determined for alloys where both the corrosion potential and pitting potential were reported, or able to be determined by the authors.

Some high-level observations from the data representation in **Figure 6a**, include: (i) it is evident that the majority of alloys (irrespective of test electrolyte) reveal a corrosion potential between ~ -500 to ~0 mV$_{SCE}$; (ii) that for a given potential, alloys present a spread of corrosion rates over a range of ~2-3 orders of magnitude; (iii) visual inspection suggests that when observing data for each electrolyte in isolation, there is a negative slope (with decreasing corrosion rate with increasing corrosion potential), which may be suggestive of corrosion being under so- called 'anodic control', (iv) many of the 'most noble' corrosion potentials were associated with HNO$_3$ as the test electrolyte (i.e. oxidizing acid), (v) many of the 'least noble' corrosion potentials were associated with NaOH as the test electrolyte (i.e. highly alkaline conditions), and (vi) the range of corrosion potentials varies over a vast range of nearly 1.8V.

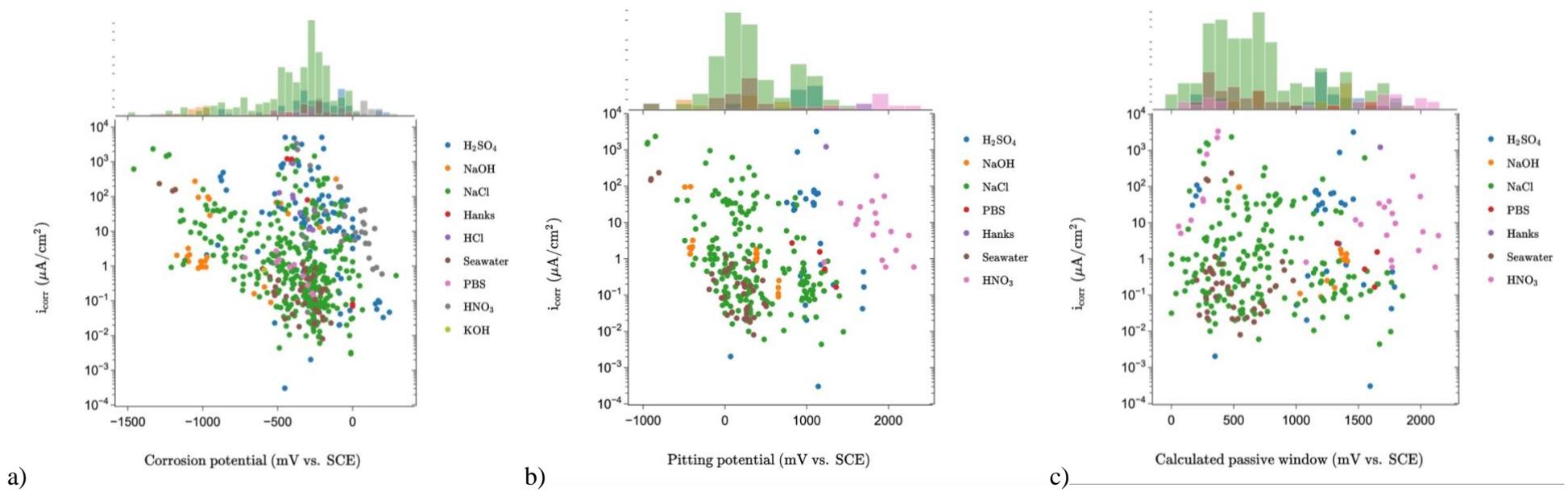

*Figure 6*. The corrosion current density data versus the corresponding a) corrosion potential, b) pitting potential, and c) calculated passive window for alloys in the MPEA corrosion database, with data disambiguated by test electrolyte. Additionally, the univariate distribution of the parameter represented by the x-axis is shown in the top margin of each plot.

Some high-level observations from the data representation in **Figure 6b**, include: (i) it is evident that there is a broad range of reported pitting potentials, with a range of >3V. For pitting potentials reported that are in excess of >1V$_{SHE}$ (or ~ 0.76V$_{SCE}$), it is likely that the reported pitting potentials are transitions to transpassive dissolution (including the possibility that at even higher potentials, there is a possibility for oxygen evolution at the test electrode); (ii) some of the highest reported pitting potentials amongst all the MPEAs reported, were realised in HNO$_3$ and H$_2$SO$_4$; (iv) many of the lowest reported pitting potentials were in Cl$^-$ containing electrolytes (NaCl and seawater), whilst noting a scarcity of reported pitting potentials in HCl.

Some high-level observations from the data representation in **Figure 6c**, include: (i) the data reveals a significant spread across the values of i$_{corr}$ and calculated passive window; (ii) there appears to be no strong correlation between these parameters (even within the confines of a specific test electrolyte).

*3.2.2. Electrolyte, elemental and structural effects*

In order to provide some further insight and nuance to the role of electrolyte beyond what is embodied in the overview plots in **Figure 6**, a representation of the role of electrolyte upon i$_{corr}$ has been provided in **Figure 7**.

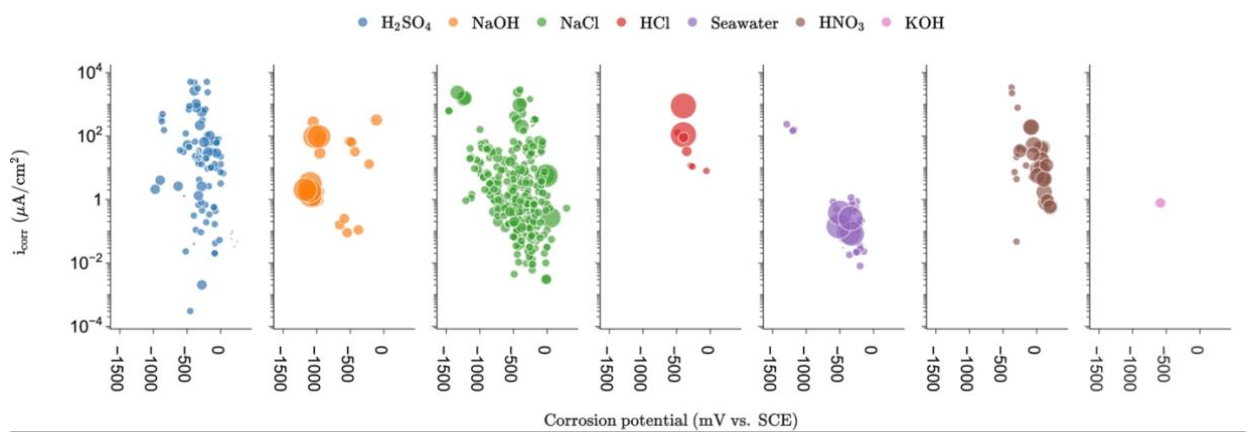

*Figure 7. The role of the test electrolyte on the reported corrosion potential and corrosion current density for alloys in the MPEA corrosion database. The size of the datapoints represents the relative concentration of the test electrolyte.*

The representation in **Figure 7** allows for an assessment of the relationship between corrosion potential and i$_{corr}$, whilst also revealing the concentration of electrolyte via the relative size of the corresponding datapoints. This representation allows for some interpretable trends, that include: (i) the highest i$_{corr}$ values correspond to testing in HCl. (ii) for tests carried out in NaOH, the lowest corrosion potentials are associated with the highest NaOH concentrations, (iii) there is a vast spread of results with less of an apparent trend for tests carried out in NaCl and H$_2$SO$_4$, which have the highest volume of associated data, and (iv) the majority (>~70% of data) for NaCl present i$_{corr}$ values lower than 1 $\mu$A/cm$^2$, whilst the majority of data for testing in H$_2$SO$_4$ present i$_{corr}$ values greater than 1 $\mu$A/cm$^2$.

In order to provide some further insight and nuance to the role of alloying element type, an overview plots have been presented in **Figure 8**. Given the complexity of MPEA compositions, and the large number of variables in their interpretation – the presentation in **Figure 8** is one attempt at isolating the role of individual elements on the corrosion rate of MPEAs in which they are present. It is noted that readers may wish to create their own representations via the use of source data in the MPEA database, which may allow relationships to be realised for elements not present in Figure 8, or the relationships between multiple elements simultaneously.

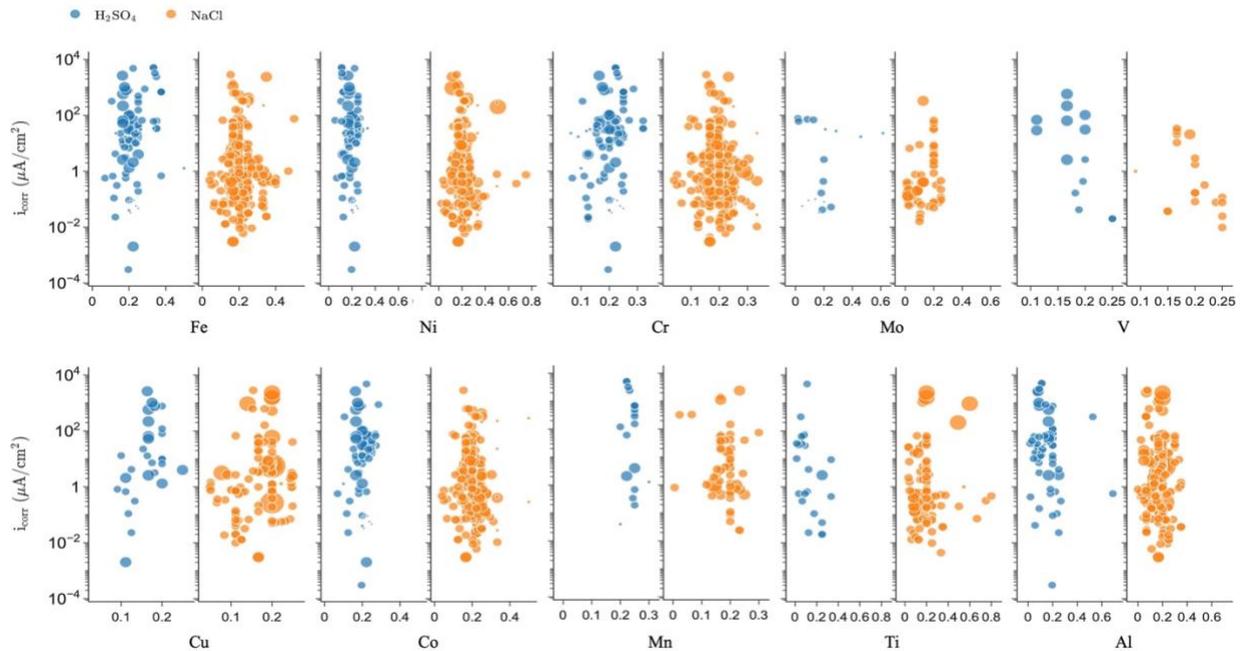

*Figure 8. The relationship between the corrosion current density and the alloy fraction of the most important alloying elements in MPEAs (represented by atomic fraction), including Fe, Ni, Cr, Mo, V, Cu, Co, Mn, Ti, and Al. The data has been presented in a manner that prioritises (and separates) the two most prevalent test electrolytes, $H_2SO_4$ and NaCl. The size of the datapoints represents the relative concentration of the test electrolyte.*

From **Figure 8**, which also provides data from testing in the two key electrolytes from corrosion studies of MPEAs, some insight include: (i) there is a broad spread of corrosion rates (6 orders of magnitude) associated with elements considered as a 'corrosion resistant elements' (in isolation), including Ni, Cr, Co and Ti; (ii) The corrosion rate of MPEAs is notably greater in $H_2SO_4$ for alloys that contain Fe, Co, or Al; (iii) the presence of Cr does not assure low rates of corrosion, (iv) although there is comparatively less data for MPEAs containing V and Mo, it appears that MPEAs containing either V or Mo typically have low rates of corrosion (lower than 1 $\mu A/cm^2$) in NaCl containing electrolytes; and (v) the lack of obvious trends arising from the presence of individual elements (irrespective of test electrolyte) reveals that the corrosion behaviour of MPEAs is principally dictated by the complex and compound effects of elemental combinations (and their proportions). This is concomitant with corrosion kinetics (and subsequent passive film chemistry) being reliant on the rate of incongruent dissolution (where incongruent dissolution is by definition, reliant on the co-presence of multiple principal elements in the alloy).

As the crystal structure is listed for each alloy in MPEA corrosion database, there is also an opportunity to assess the role of alloy crystal structure via an overview plot – which has been done in **Figure 9**. It is revealed that from **Figure 9**, there is no obvious trends that one crystal structure is beneficial (or conversely, detrimental) relative to the other. The 'other' category in **Figure 9** includes alloys that may contain an HCP phase, an intermetallic phase, an ordered phase (such as B2), or distinct minor phases (i.e. sigma, Laves, etc.).

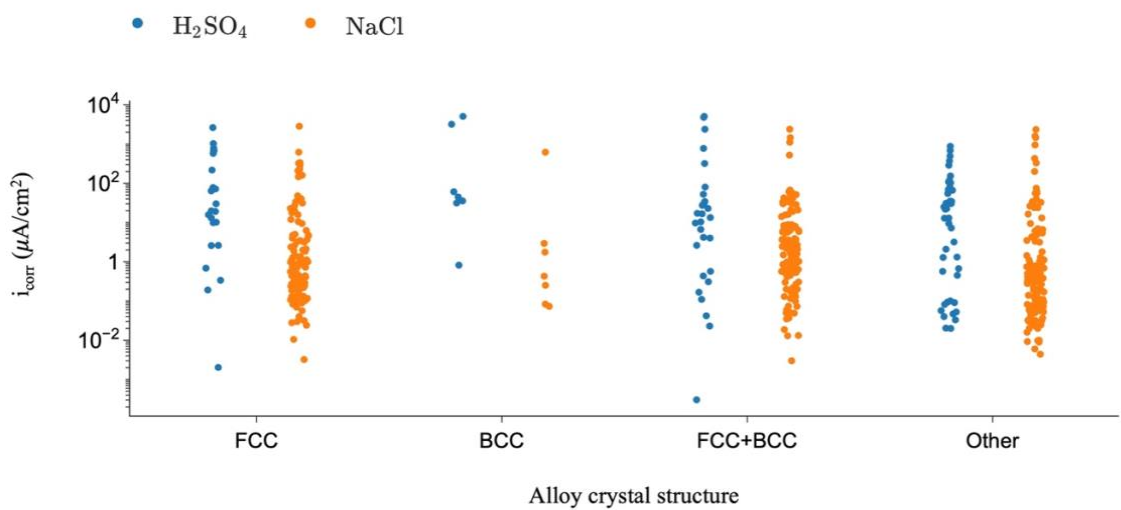

*Figure 9. MPEA crystal structure reported versus the corresponding corrosion current density data for all alloys in the MPEA corrosion database. Data is presented for the two most prevalent test electrolytes, $H_2SO_4$ and NaCl.*

### 3.3 Towards data science and machine learning

To visualise the MPEA corrosion data of high-dimensionality, in a lower-dimensional space, the t-SNE dimension reduction algorithm was used to generate two-dimensional representations consisting of t-SNE component 1 and t-SNE component 2. The practice of dimension reduction is common in data science, the process of which has been defined in several seminal works [60-63] and examples in the context of materials properties (specifically for alloy materials) can be readily found [12, 64-69]. The t-SNE notation refers to t-distributed stochastic neighbour embedding that reduces the dimensionality of data, whilst preserving relationships between neighbouring data points. The so-called "t-distributed" and "neighbour embedding" features measure the similarities between data points in high-dimensional space, and the corresponding low-dimensional embedding space. This enables t-SNE to capture complex nonlinear relationships between data points. The "stochastic" feature of data representation using t-SNE dimension reduction algorithm involves a degree of randomness for the optimisation process to avoid local minima.

However, the utilisation of lower-dimensional space is not only useful for visual inspection, but also alloys the ability to apply data sorting and analysis, in the curation of plots in lower-dimensional space . To this end, a 'clustering' analysis known as 'density based spatial clustering of applications with noise' (or, DBSCAN) [70-74] was performed. The reason why clustering analysis is important in data science related to materials, is because it provides a computationally robust and algorithmically-informed insight to the human researcher regarding trends in data of high-dimensionality that are otherwise not evident. A large database will notionally have some links between portions of data, meaning for example that subsequent treatment of data may be best carried out by a treatment of subsets of data.

The BDSCAN method, like most algorithmic clustering schemes, is an unsupervised machine learning (ML) method. The BDSCAN method was specifically selected for application to the MPEA corrosion database, as it is robust to noise and outliers [75]. The purpose of a clustering analysis of complex datasets, is to identify so called 'clusters' based on the density of the data points (i.e. the number of the data points (MinPts)) within a specified radius around each data point denoted as epsilon ($\varepsilon$), whereby $\varepsilon$ and MinPts are user specified parameters. If the number of data points within $\varepsilon$ exceeds MinPts, a so-called 'dense' region (termed a cluster) is formed. When applying this approach to the data in the MPEA corrosion database, eight distinct clusters were identified and reported in **Figure 10**, showing similarities or patterns in the data. The input data for the DBSCAN includes MPEA crystal structure, test electrolyte, chemical composition of MPEA, corrosion potential, and corrosion current density. As clustering is an unsupervised ML method, the corrosion properties were not pre-defined as labels here, meaning that the algorithm groups similar data points together based solely on their inherent structure or similarity in the feature space, without any prior knowledge of class labels.

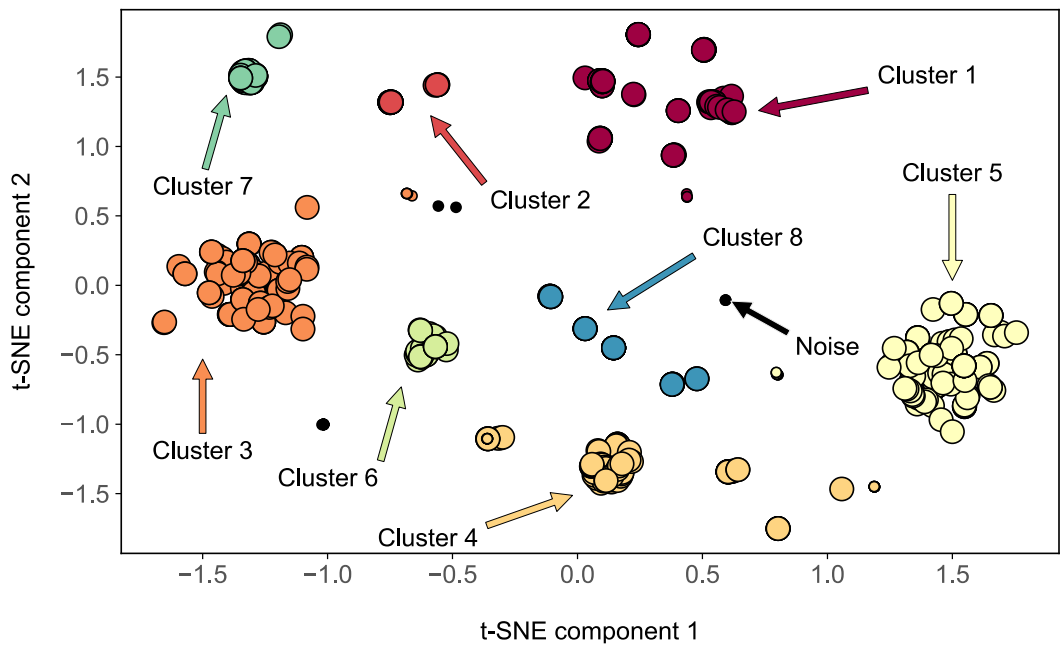

*Figure 10. DBSCAN clustering analysis in combination with t-SNE dimension reduction algorithm revealing eight distinct clusters separated based on the similarities in the MPEA corrosion database.*

From an inspection of the clustering analysis that is represented by **Figure 10**, the clustering algorithms have made the following 'classes' represented in **Table 4**.

*Table 4. Characteristics of BDSCAN clustering algorithm applied to the MPEA corrosion database.*

| Cluster # | Cluster characteristic | Comment |
|---|---|---|
| 1 | Inclusive of specimens tested in $H_2SO_4$ | Count of 105 |
| 2 | Inclusive of specimens tested in NaOH | Count of 26 |
| 3 | Inclusive of specimens tested in NaCl, HCl and KOH | Count of 112 |
| 4 | Inclusive of specimens tested in NaCl, Hank's solution and $HNO_3$ | Count of 97 |
| 5 | Inclusive of specimens tested in only NaCl | Count of 120 |
| 6 | Inclusive of specimens tested in high concentrations of NaCl | Count of 35 |
| 7 | Inclusive of specimens tested in seawater | Count of 30 |
| 8 | Inclusive of tested in phosphate buffered saline, $HNO_3$, NaCl and seawater | Count of 55 |

It is evident from **Table 4,** together with a manual inspection of the associated MPEAs within each class, that the DBSCAN classification algorithm created clusters based on subsets of the data that were at the high level, grouped by test electrolyte. The clustering was however nuanced, in that some clusters only had entries of alloys tested in a single electrolyte, but other clusters combined electrolytes, suggestive of relationships in each cluster. It is noted, that for many clusters, there was a diversity of MPEAs (i.e. inclusive of MPEAs that may have included or excluded certain alloying elements, along with a variety of alloy crystal structures, for example). Therefore, whilst the BDSCAN clustering was not discriminating classes based on MPEA composition alone (although, composition will impact corrosion rate), it is noted that clustering analysis does not seek to classify groups of data on output functions, but rather input functions, which has been discussed in some detail by [76] and co-workers [77].

Because the data in Table 4 reveals that clusters at the highest level are influenced by test electrolyte, it is therefore relevant to extract some more detail from the clusters, by an inspection of trends in MPEA composition as they relate to their cluster – as represented in **Table 5**.

Although the BDSCAN clustering algorithm did not discriminate classes based on MPEA composition, the trends evident from inspection of Table 5 are concisely described. Cluster 1 was the only cluster that contains any Be, Ga, and La, while this cluster did not contain any Mg, Hf, Sn, or Zn. In addition, Cluster 1, with 105 data points, includes a rich region of Al with the maximum content of 0.69 atomic fraction. This same cluster also had a maximum content for B at 0.15 atomic fraction, the maximum content of Fe at 0.50 atomic fraction, the maximum content of Mo at 0.61 atomic fraction, and the maximum content of Ta at 0.20 atomic fraction. This cluster shares similarities with Cluster 4 which also contains Al and Ta rich MPEAs; whilst also exhibiting some similarities to Cluster 3 in terms of containing MPEAs rich-B and rich-Fe. Conversely, Cluster 3 presented the maximum contents (in atomic fraction) for Co (0.5), and Ni (0.75). Meanwhile, Cluster 4 presented the maximum content (in atomic fraction) for Hf (0.27).

Cluster 2 was the only cluster in which all alloys therein contained Co and Fe in their chemical composition. This cluster shared similarities with Cluster 7 and Cluster 8, with no Mn, no W, and no B, along with a comparatively low maximum content of Cr at 0.18 atomic fraction. Both Cluster 2 and Cluster 7 do not contain Nb and Zr, making them distinct from Cluster 8. While Cluster 7 is Zr-free, Cluster 8 has the maximum content of Zr with 0.37 atomic fraction. As a result, it can be noted that the BDSCAN clustering algorithm is capable of making distinct – and meaningful - classifications from the MPEA database. Such class-based analysis of databases, is an important precursor to subsequent supervised machine learning operations, in order to allow machine learning to be 'interpretable' [78, 79]. Cluster 5 was the largest cluster with 120 data points, containing the maximum Ti content, rather high Ni content (0.66 atomic fraction) and the highest Si content (0.20 atomic fraction). Cluster 6 was the only cluster to contain any Mg or any Zn, whilst sharing similarities with Cluster 5 in regard to containing similar ranges for Ni, Si, and Ti. The two clusters, Cluster 6 and Cluster 7, were the only V-free clusters, whilst Cluster 7 was the only cluster with no Ti and with the maximum content of Fe (0.53 atomic fraction).

*Table 5. Characteristics from application of the BDSCAN clustering algorithm applied to the MPEA corrosion database, presented as chemical composition. The 'range' represented the alloy fraction (in atomic fraction).*

| Element | Range | Cluster number | | | | | | | |
|---|---|---|---|---|---|---|---|---|---|
| | | 1 | 2 | 3 | 4 | 5 | 6 | 7 | 8 |
| **Al** | min. | 0.00 | 0.00 | 0.00 | 0.00 | 0.00 | 0.00 | 0.02 | 0.00 |
| | max. | 0.69 | 0.29 | 0.36 | 0.55 | 0.35 | 0.35 | 0.29 | 0.29 |
| **B** | min. | 0.00 | 0.00 | 0.00 | 0.00 | 0.00 | 0.00 | 0.00 | 0.00 |
| | max. | 0.15 | 0.02 | 0.15 | 0.00 | 0.00 | 0.00 | 0.00 | 0.00 |
| **Be** | min. | 0.00 | 0.00 | 0.00 | 0.00 | 0.00 | 0.00 | 0.00 | 0.00 |
| | max. | 0.25 | 0.00 | 0.00 | 0.00 | 0.00 | 0.00 | 0.00 | 0.00 |
| **Co** | min. | 0.00 | 0.12 | 0.00 | 0.00 | 0.00 | 0.00 | 0.00 | 0.00 |
| | max. | 0.29 | 0.27 | 0.50 | 0.25 | 0.30 | 0.33 | 0.29 | 0.20 |
| **Cr** | min. | 0.00 | 0.00 | 0.00 | 0.00 | 0.00 | 0.00 | 0.05 | 0.00 |
| | max. | 0.32 | 0.18 | 0.33 | 0.25 | 0.30 | 0.33 | 0.20 | 0.17 |
| **Cu** | min. | 0.00 | 0.00 | 0.00 | 0.00 | 0.00 | 0.00 | 0.00 | 0.00 |
| | max. | 0.25 | 0.25 | 0.25 | 0.24 | 0.25 | 0.20 | 0.05 | 0.20 |
| **Fe** | min. | 0.00 | 0.12 | 0.00 | 0.00 | 0.00 | 0.00 | 0.00 | 0.00 |
| | max. | 0.50 | 0.25 | 0.50 | 0.33 | 0.38 | 0.33 | 0.53 | 0.33 |
| **Ga** | min. | 0.00 | 0.00 | 0.00 | 0.00 | 0.00 | 0.00 | 0.00 | 0.00 |
| | max. | 0.25 | 0.00 | 0.00 | 0.00 | 0.00 | 0.00 | 0.00 | 0.00 |
| **Hf** | min. | 0.00 | 0.00 | 0.00 | 0.00 | 0.00 | 0.00 | 0.00 | 0.00 |
| | max. | 0.00 | 0.00 | 0.00 | 0.28 | 0.00 | 0.20 | 0.00 | 0.25 |
| **La** | min. | 0.00 | 0.00 | 0.00 | 0.00 | 0.00 | 0.00 | 0.00 | 0.00 |
| | max. | 0.02 | 0.00 | 0.00 | 0.00 | 0.00 | 0.00 | 0.00 | 0.00 |
| **Mg** | min. | 0.00 | 0.00 | 0.00 | 0.00 | 0.00 | 0.00 | 0.00 | 0.00 |
| | max. | 0.00 | 0.00 | 0.00 | 0.00 | 0.00 | 0.25 | 0.00 | 0.00 |
| **Mn** | min. | 0.00 | 0.00 | 0.00 | 0.00 | 0.00 | 0.00 | 0.00 | 0.00 |
| | max. | 0.30 | 0.00 | 0.30 | 0.22 | 0.25 | 0.17 | 0.00 | 0.00 |
| **Mo** | min. | 0.00 | 0.00 | 0.00 | 0.00 | 0.00 | 0.00 | 0.00 | 0.00 |
| | max. | 0.62 | 0.13 | 0.13 | 0.25 | 0.20 | 0.13 | 0.00 | 0.01 |
| **Nb** | min. | 0.00 | 0.00 | 0.00 | 0.00 | 0.00 | 0.00 | 0.00 | 0.00 |
| | max. | 0.33 | 0.00 | 0.17 | 0.33 | 0.20 | 0.25 | 0.00 | 0.25 |
| **Ni** | min. | 0.00 | 0.00 | 0.00 | 0.00 | 0.00 | 0.00 | 0.20 | 0.00 |
| | max. | 0.33 | 0.27 | 0.75 | 0.26 | 0.67 | 0.51 | 0.35 | 0.20 |
| **Si** | min. | 0.00 | 0.00 | 0.00 | 0.00 | 0.00 | 0.00 | 0.00 | 0.00 |
| | max. | 0.02 | 0.00 | 0.00 | 0.20 | 0.20 | 0.15 | 0.00 | 0.00 |
| **Sn** | min. | 0.00 | 0.00 | 0.00 | 0.00 | 0.00 | 0.00 | 0.00 | 0.00 |
| | max. | 0.00 | 0.02 | 0.02 | 0.00 | 0.02 | 0.04 | 0.00 | 0.00 |
| **Ta** | min. | 0.00 | 0.00 | 0.00 | 0.00 | 0.00 | 0.00 | 0.00 | 0.00 |
| | max. | 0.20 | 0.00 | 0.00 | 0.25 | 0.00 | 0.00 | 0.00 | 0.00 |
| **Ti** | min. | 0.00 | 0.00 | 0.00 | 0.00 | 0.00 | 0.00 | 0.00 | 0.00 |
| | max. | 0.33 | 0.14 | 0.55 | 0.38 | 0.80 | 0.60 | 0.00 | 0.45 |
| **V** | min. | 0.00 | 0.00 | 0.00 | 0.00 | 0.00 | 0.00 | 0.00 | 0.00 |
| | max. | 0.25 | 0.17 | 0.17 | 0.36 | 0.20 | 0.00 | 0.00 | 0.29 |
| **W** | min. | 0.00 | 0.00 | 0.00 | 0.00 | 0.00 | 0.00 | 0.00 | 0.00 |
| | max. | 0.20 | 0.00 | 0.20 | 0.20 | 0.20 | 0.00 | 0.00 | 0.00 |
| **Y** | min. | 0.00 | 0.00 | 0.00 | 0.00 | 0.00 | 0.00 | 0.00 | 0.00 |
| | max. | 0.00 | 0.00 | 0.00 | 0.00 | 0.00 | 0.20 | 0.00 | 0.00 |
| **Zn** | min. | 0.00 | 0.00 | 0.00 | 0.00 | 0.00 | 0.00 | 0.00 | 0.00 |
| | max. | 0.00 | 0.00 | 0.00 | 0.00 | 0.00 | 0.30 | 0.00 | 0.00 |
| **Zr** | min. | 0.00 | 0.00 | 0.00 | 0.00 | 0.00 | 0.00 | 0.00 | 0.00 |
| | max. | 0.17 | 0.00 | 0.00 | 0.28 | 0.25 | 0.20 | 0.00 | 0.37 |

In order to further analyse the characteristics of the BDSCAN determined Clusters, a combination of information from **Figure 9**, **Table 4** and **Table 5** was curated. This combines the corrosion current density depicted as a function of Clusters, together with information regarding MPEA crystal structure, test electrolyte and Cluster characteristics, shown in **Figure 11**.

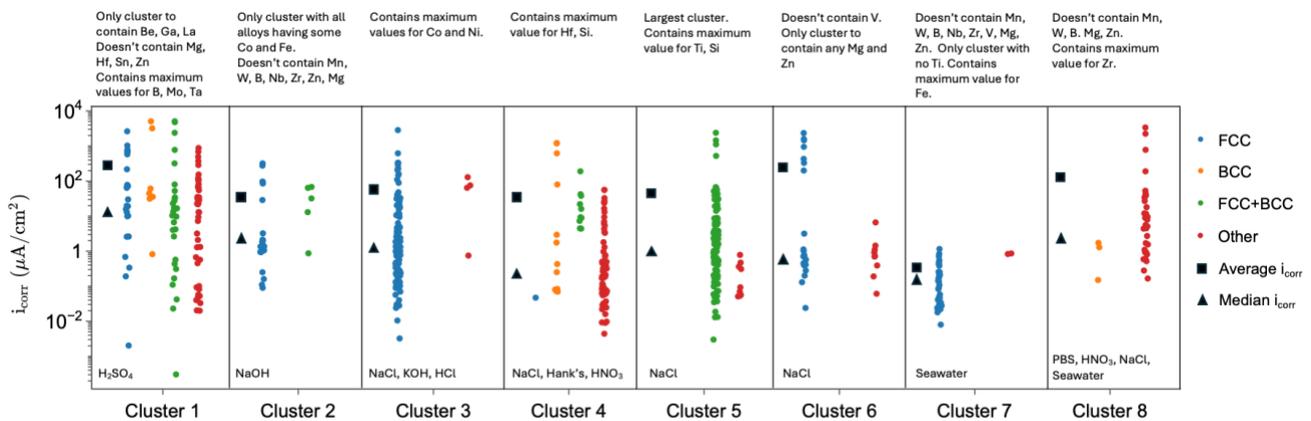

*Figure 11. Eight distinct DBSCAN clusters versus the corresponding corrosion current density and supporting information including the four broad categories of MPEA crystal structure, FCC, BCC, FCC+BCC, Other.*

Figure 11 is insightful on the basis that there are variations in the median $i_{corr}$ in each cluster, varying from 0.14 $\mu A/cm^2$ for Cluster 7, to 21.87 $\mu A/cm^2$ for Cluster 1. It is also visually apparent that entries from MPEAs with specific crystal structures dominate each cluster. For example, Cluster 5 did not contain any single phase FCC or BCC MPEAs, along with several other discernible trends evident across Figure 11. The 'sorting' evident in Figure 11 is considered an important outcome in this review, as it provides an interpretable classification for various MPEAs in diverse test environments.

Whilst not elaborated in the present review, the authors note that another clustering algorithm used in combination with the t-SNE dimension reduction algorithm was explored, known as the K-Means method [80-83]. The K-Means method starts by randomly placing K centroids in the data space, followed by assigning each data point to the nearest centroid (typically calculated using Euclidean distance). Such an alternative clustering approach was also effective – but not elaborated because it did not offer any distinct advantages to the BDSCAN approach, and no additional value for the purposes of this review. Alternative approaches are however important to mention, should readers wish to apply their own data science approaches to MPEA database. Cluster analysis applied to a dataset is beneficial in identifying the underlying structure and patterns in the data. Thus, by assessing test electrolytes, MPEA compositions, and alloy crystal structures together, clustering analysis facilitated the identification of commonalities and differences between different MPEAs that have undergone corrosion testing. Additionally, Cluster analysis has provided valuable insights into the corrosion performance of MPEAs, and which input features may drive such properties. Whilst not the purpose of this review, harnessing such information is valuable for optimising MPEAs for corrosion performance (including in specific electrolytes.

Furthermore, Cluster analysis enables the identification of outliers within a given dataset. Such outliers could represent compositions with unique or atypical properties; and by identifying such outliers, researchers may focus attention on potentially significant compositions that may warrant experimental or computational analysis. Moving forward, several logical next steps can be considered in terms of machine learning. Firstly, integrating Clustering results with supervised machine learning techniques could enhance the predictive capability of property prediction models for MPEAs. More precisely, clustering assignments could be used as features in supervised learning algorithms to predict alloy properties based on composition and crystal structure. This integration of unsupervised and supervised learning approaches can lead to more accurate and comprehensive models for alloy design and optimisation. The application of advanced data science / machine learning methods such as deep learning can offer new avenues for analysing and understanding MPEAs, whilst such methods may also further help uncover complex patterns or nonlinear relationships within the data that may not be fully determined through unsupervised analysis methods.

## 4. Summary and future prospects

This concise review has sought to provide a high-level overview of the present critical understanding in the corrosion (and passivation) of multi-principal element alloys. The review initially covered advances in the open literature with the following findings.

- Corrosion of MPEAs was dominated by evidence of incongruent dissolution in essentially all cases. This means that MPEAs dissolve in a manner that is not harmonious with the bulk alloy stoichiometry.
- Furthermore, the nature of incongruent dissolution was deemed to be complex, in that there are instances where small alterations in alloy composition led to significant changes in the extent (proportion-wise) of incongruent dissolution; and other cases where in one alloy a preferentially dissolving element (i.e. Cr) may be the slowest to dissolve in another alloy of differing composition.
- The composition of surface / passive films upon MPEAs were also determined to be highly complex. This includes the following characteristics:
    - Compositions that differ in the proportion of elements in the base alloy (as a result of the aforementioned incongruent dissolution)
    - A complex bi-layer structure that varies in composition from the outer surface to the bulk metallic (alloy) substrate.
    - The presence of unoxidized metal in the surface film on some MPEAs, at times characterized by a non-trivial proportion of unoxidized metal.
    - The possibility for complex heterogenous compounds in the surface film, that may include various hydroxides and spinel phases [84].
- The above points emphasise to the unique electrochemical facets of MPEA corrosion. There is, as a result, no universal statement that wholly captures the behaviour of MPEAs as a class of alloys. However, the prevalence of incongruent dissolution and complex oxides are features generally associated with MPEAs. On this basis, there is important prospects emerging to be able to capture such complexity in models of passivity. Some works regarding MPEAs to date have sought to align the passivation (and film growth) kinetics of MPEAs with the Point Defect Model [85], with some success [54]. On this basis, emerging empirical studies such as those of Wang and co-workers [55, 56] provide significant prospect for MPEAs to be interpreted in the context of a passivity framework.
- Most recently, during preparation of this review, there has also been a demonstration of using high throughput density functional theory (DFT) calculations for the assessment of electrochemical performance of MPEAs, in a study by Yuwono and co-workers [86]. Such work is indispensable in providing a fundamental mechanistic basis to the dissolution kinetics (including incongruent dissolution) for MPEAs – and sets a strong platform for multidisciplinary studies moving forward, which is a very important area for future work. Such recent work is potentially revolutionary, on the basis that prior experimental results are not essential as input parameters. It is noted that there have been additionally a small (but very important) set of studies that have employed DFT in the study of MPEAs, from the work of Lu et al. [87], through to more recent studies [88, 89], including into 2024 [90-92].

A key aspect of the present review, was to compile and present a comprehensive database of corrosion properties of MPEAs, reported in the open literature. To this end, the largest database of such information to date was prepared and provided. From that database, a number of higher-level observations were made, all of which added to the notion that a universal understanding of corrosion of MPEAs is complex, and corrosion of each MPEA is therefore nuanced (whilst occurring via generally similar mechanisms). From the MPEA corrosion database, a small subset of alloys that had the highest, and lowest, reported corrosion current densities are shown in **Table 6**.

It can be observed from Table 6 that the alloys with the highest reported corrosion rates were notionally tested in 0.5 M $H_2SO_4$, however contained compositions inclusive of numerous elements that are notionally corrosion resistant. The alloys with the lowest reported corrosion rates were notionally tested in NaCl, however they were inclusive of elements that were not significant deviations from alloys with the highest reported corrosion rates. This highlights the sensitivity of MPEAs to minor changes in composition, and indeed to an (expected) influence from test electrolyte.

*Table 6. The MPEAs corresponding to the highest and lowest of the reported rates of corrosion current density in the MPEA corrosion database compiled for this review.*

| Alloy name | Phases present | Test electrolyte | Corrosion potential ($mV_{SCE}$) | Pitting potential ($mV_{SCE}$) | $i_{corr}$ ($\mu A/cm^2$) |
|---|---|---|---|---|---|
| *Highest reported corrosion rates* | | | | | |
| $Al_{0.5}CrFe_{1.5}MnNi_{0.5}$ | FCC+BCC | 0.5 M $H_2SO_4$ | -206 | | 5080 |
| $AlCoCrFeNiTi_{0.5}$ | FCC+BCC | 0.5 M $H_2SO_4$ | -390 | | 4800 |
| AlCoCuFeNiTi | FCC+A2/B2+Laves | 0.5 M $H_2SO_4$ | -374 | | 3379 |
| $Al_{0.4}CrFe_{1.5}MnNi_{0.5}$ | BCC | 0.5 M $H_2SO_4$ | -340 | 1120 | 3200 |
| $Al_{0.5}CoCrCuFeNiB$ | FCC | 0.6 M NaCl | -403 | | 2850 |
| *Lowest reported corrosion rates* | | | | | |
| $Al_{0.5}CoCrFeNi$ | Not reported | 0.6 M NaCl | -220 | 480 | 0.006 |
| $TiZr_{0.5}NbCr_{0.5}$ | BCC+Laves | 0.6 M NaCl | -489 | 1180 | 0.00441 |
| AlCrFeNiCoCu | FCC | 1M NaCl | -12 | | 0.00323 |
| AlCrFeNiCoCu | FCC + BCC | 1M NaCl | -12 | | 0.003 |
| $Cu_{0.5}CoCrFeNi$ | FCC | 1M $H_2SO_4$ | -280 | 70 | 0.00203 |

The data science aspects of the review were confined to reviewing data, as opposed to an original study that employs supervised machine learning for handling the data in the MPEA corrosion database. However, there is significant potential in follow on / future works that apply supervised machine learning to the MPEA corrosion database presented in this review. This is because the high dimensionality of the data, and its complexity in human interpretation, presents an ideal use scenario for machine learning. Some obvious facets of a greater understanding of corrosion in the context of future MPEA design include [93]sation for corrosion in lightweight MPEAs, or MPEAs produced from recycled materials/alloys. The present review has not explored the high temperature oxidation of MPEAs, which is an intense area of study (owing to the prospect of refractory MPEAs/HEAs for demanding applications).

**Acknowledgements**

The technical assistance of following individuals in the curation (including data mining) of the database presented herein is gratefully acknowledged: Theo Darmawan, Ruijia Tan, Ninad Bhat, Himadri Shekhar Mondal and Shujing Zhao. Financial support from the Office of Naval Research under the contract ONR: N00014-17-1-2807 with Dr. David Shifler and Dr. Clint Novotny as program officers is gratefully acknowledged.

**Competing interests**

The authors declare no competing or conflicts of interests.